%% file: main.tex
\documentclass[
 aps, pra,
 amsmath,amssymb,
 11pt,
 final,
tightenlines,
 twoside,
 twocolumn,
 nofloats,
nofootinbib,
 superscriptaddress,
showkeys,
showkeywords,
 ]
{revtex4-2}

\usepackage[T1]{fontenc}
\usepackage[utf8x]{inputenc}
\usepackage[english]{babel}
\usepackage{graphicx}
\usepackage{dcolumn}
\usepackage{bm}

\input{maik.rty}

\setcitestyle{authoryear,round}
\input{sao_cmd_author.tex}

\begin{document}

\selectlanguage{english}


\title{The Metron Project --- I. The Metron Project Science Program}
\author{V. K. Dubrovich}\thanks{e-mail: dvk47@mail.ru}
\affiliation{St. Petersburg Branch of Special Astrophysical Observatory of Russian Academy of Sciences, \\65 Pulkovskoye Shosse, St Petersburg, 196140 Russia}
\author{S. I. Grachev}\thanks{e-mail: s.grachev@spbu.ru}
\affiliation{St. Petersburg State University, St. Petersburg, 199034 Russia}
\author{Yu. N. Eroshenko}\thanks{e-mail: eroshenko@inr.ac.ru}
\affiliation{Institute for Nuclear Research of Russian Academy of Sciences, Prospekt 60-Letiya Oktyabrya, 7a, Moscow, 117312 Russia}
\author{S.~I.~Shirokov}\thanks{e-mail: lakronous@mail.ru, Corresponding author}
\affiliation{St. Petersburg Branch of Special Astrophysical Observatory of Russian Academy of Sciences, \\65 Pulkovskoye Shosse, St Petersburg, 196140 Russia}
\author{and G. G. Valyavin}\thanks{e-mail: gvalyavin@gmail.com}
\affiliation{Special Astrophysical Observatory of Russian Academy of Sciences, Nizhny Arkhyz, \\Zelenchuksky District, Karachay-Cherkess Republic, 369167 Russia}

\date{\today}

\begin{abstract}
In this paper, the first in a series of four articles, the scientific goals of the Metron project are highlighted, and the characteristics of the cosmic objects available for study within its framework are provided. The Metron interferometer radio telescope should include arrays of meter-range dipole antennas placed on Earth, in outer space or on the far side of the Moon (or a combination of these options). Working in the meter range will enable the study of the so-called cosmological epoch of ``Dark Ages'', which is challenging to observe but highly interesting for understanding the origin of the first stars, galaxies, and black holes, as well as for the search for new cosmological objects and processes. One possibility is to search for absorption in the 21 cm line within extended halos around early protogalaxies and supermassive primordial black holes, whose existence is predicted in a number of models. Another goal of Metron may be to clarify the anomalous absorption in the 21 cm line previously detected by the EDGES telescopes and to observe radio emissions from stars' and exoplanets' magnetospheres. The Metron project aims to achieve unprecedented resolution for the meter range, which is expected to yield new world-class scientific results. Meter-range antennas and receivers are relatively simple and inexpensive, and the construction of interferometric arrays from them can be accomplished in a relatively short period of time. 
\end{abstract}

\maketitle 

\section{Introduction}

From the time of hydrogen recombination to the appearance of the first stars, the Universe had no sources of radiation, hence this period of cosmological history is referred to as the ``Dark Ages''. Researching the Dark Ages is of great interest to cosmology, since at the end of this epoch, stars, galaxies, and supermassive black holes begin to form. The unsolved mystery is the origin of supermassive black holes in early epochs. Quasar observations have shown that black holes with masses $\geq10^9M_\odot$ already existed at redshifts $z>7$. The presence of such massive black holes is difficult to explain through conventional astrophysical processes and requires initial objects with masses $\sim(10^4-10^5)M_\odot$. These objects could be primordial black holes (PBHs), which can form through various mechanisms
~\citep{ZelNov66,Haw71,Car75,KhlPol0,BerKuzTka83,RubSakKhl01,DolSil93}.

The James Webb Space Telescope recently unveiled unexpected data on galaxies at the end of the Dark Ages. It turns out that sufficiently large structured galaxies started forming even before the Universe's reionization (see, for example, \citet{Foretal23}). Another issue is that some of the early galaxies exhibit a significant deficit of dark matter~\citep{Cometal23}. This puzzle also requires an explanation. Another problem that remains unsolved is the anomalous absorption in the 21 cm line of neutral hydrogen, detected by the EDGES telescopes \citep{BowRogMon18}.

To study the Dark Ages, we propose creating several arrays of meter-range antennas operating as interferometers. Each array should consist of a square kilometer-sized platform with multiple antennas. Ground-based meter-range antenna arrays are relatively inexpensive systems and can be built in a relatively short period (on the order of a few years). Even with ground-based telescopes, it will be possible to obtain fundamentally new results, potentially discovering new cosmological objects discussed further in this article. Launching such arrays into space opens unprecedented opportunities that have not yet been utilized in the global radio astronomy community.

More detailed information on the technical aspects of implementing this project and the characteristics of the planned meter-range telescopes will be described in subsequent articles. In this article, we will discuss the potential scientific objectives of this project.

The Dark Ages era is highly promising for the search for rare cosmological objects in the radio frequency range due to the scarcity of stars, galaxies, and accreting black holes before reionization, resulting in a lack of strong local obstacles and backgrounds. At redshifts $z>15$, there is a significant amount of neutral hydrogen, and at large $z$ (but $z\ll 150$), there is a substantial temperature difference between matter and the cosmic microwave background, allowing for the observation of absorption of the relic radiation in the 21 cm line~\citep{PriLoe12}. Based on this, we consider redshifts near $z\sim20$ to be the most optimal for observing rare massive objects, although there is in principle the possibility to observe objects throughout the entire epoch of $10<z<150$.

High-mass PBHs at redshifts $z>20$ are expected to be surrounded by halos consisting of dark matter and baryonic gas \citep{DokEro01,DokEro03}. In the study \citet{DubrovichGla12}, it is mentioned that the gas near PBHs may exhibit anomalies in chemical composition compared to the predictions of primordial nucleosynthesis theory in a uniform Universe. Such massive and concentrated objects in the early Universe are referred to as ``cosmological dinosaurs''. In the work \citet{DubrovichGraEro21}, it is shown that while significant energy release may occur in the central region of the halo due to accretion onto PBHs, hydrogen remains neutral in the outer regions of the halo and can be observed absorbing in the 21 cm line.

In the modern era, ``cosmological dinosaurs'' are expected to represent a rare class of dense spheroidal galaxies \citep{SuLiFen23}. The distance to the nearest one of them may be very large, exceeding 50 Mpc, which makes their search challenging with optical telescopes \citep{DubrovichGraEro21}. Conversely, with the Metron telescopes, it will be possible to observe numerous such objects during the Dark Ages.


\section{Early protogalaxies}

This project proposes the creation of an array of radio telescopes that will be capable of detecting extended objects during the Dark Ages, even if these objects do not have their own energy emission. Their observation will be conducted in the 21 cm absorption line, with the aim of observing individual objects rather than averaging, as done, for example, by the EDGES telescopes. This will be possible if the angular scale of these objects is sufficiently large. Examples of such objects include early protogalaxies. The angular resolution of the planned Metron telescopes will be sufficient for these observations within a reasonable range of parameters for the objects predicted by the models described below.

The observation of absorption lines is critically dependent on non-equilibrium processes in the gas \citep{Dubrovich77,Beretal93,DubrovichLip95,VarshHers77}. In particular, in the study \citet{VarshHers77}, it is demonstrated that the temperature difference between baryons and the relic radiation leads to distortions in the spectrum during the energy transfer from photons to hydrogen atoms in the 21 cm hydrogen line due to inelastic collisions. The gas motion state, especially the velocity distribution along the line of sight, also plays a crucial role in the depth of absorption. Strong absorption occurs when the gradient of the radial velocity becomes zero \citep{Zel78}.
In the works \citet{Dubrovich18, DubrovichGra19}, the effects related to significant inhomogeneities in matter and the presence of specific radial velocities are considered. Based on this approach, absorption on the periphery of extended halos around PBHs, where an absorption ring is formed, was calculated in \citet{DubrovichGraEro21}. Subsequently, this type of absorption is referred to as the DGE effect (Dubrovich, Grachev, Eroshenko). A similar absorption ring may also exist around protogalaxies, the existence of which has recently been demonstrated in observations by the James Webb Space Telescope at around $z\sim16$. Observing these protogalaxies with Metron would serve as an important complement to the observations with the James Webb Space Telescope and would advance towards even higher redshifts.

Observations of the cosmic microwave background fluctuations and galaxy distributions indicate that the initial cosmological density perturbations have a Gaussian distribution with high accuracy. Most abundant are the root-mean-square fluctuations and fluctuations of approximately $2\sigma$, which gave rise to present-day massive galaxies at redshifts around $z\sim1$. In a Gaussian distribution, there should also exist rarer objects, including those with masses of around $10^{10}-10^{12}M_\odot$, which formed during the Dark Ages at $z\geq10$. However, in the case of a typical spectrum of perturbations, the number of these objects at $z\geq10$ should be very small.
However, the James Webb Space Telescope has discovered the presence of massive galaxies at $z\geq10$, which were expected to form even earlier, around $z\sim16$ \citep{Naietal22, Donetal22}, although this inference is not yet indirect and is not based on direct spectroscopic data. Modifications to the spectrum of perturbations have been proposed to explain the existence of early galaxies \citep{PadLoe23}. Even independently of the theoretical explanation, the observational fact of the existence of massive protogalaxies during the Dark Ages at $z\sim16$ provides a unique opportunity for their observation through absorption in the 21 cm line.

Early protogalaxies are expected to be dense halos of dark matter. If the inner region of these galaxies has already virialized, the outer regions continue to expand, following the Hubble flow. There exists a turnaround radius where the transition from expansion to compression occurs. It is at this radius that absorption in the 21 cm line of neutral hydrogen is expected to occur \citep{VasShc12}, similar to what happens in the halo around PBHs \citep{DubrovichGraEro21}. We propose that the Metron project will be able to observe numerous such objects, providing valuable information about the distribution of high-amplitude density perturbations (the tail of the Gaussian distribution or deviations from the Gaussian law). Additionally, these observations will provide information about the state of cosmic gas during the Dark Ages, including temperature, ionization level, and more.


\section{Cosmological dinosaurs}

Another possibility is the formation of objects from dark matter and baryonic matter around supermassive PBHs, if such PBHs exist in the Universe. Current constraints on the possible abundance of PBHs in the Universe in various mass ranges are provided in the review \citet{Caretal20}. These constraints are given as upper limits on the fraction $f_{\rm PBH}$ of PBHs in the composition of dark matter. In the mass range of interest, $10^9-10^{10}M_\odot$, there are tight constraints on $\mu$-disturbances of the relic radiation associated with the dissipation of Gaussian perturbations (the Silk mechanism) that could have given rise to PBHs. However, these constraints are only applicable in models where PBHs are formed from adiabatic density perturbations, while for the formation of PBHs of such large masses through other mechanisms, they can be quite numerous. Dynamical constraints allow for values of $f_{\rm PBH}\leq10^{-5}$ \citep{Caretal20}.
The work \citet{CarKuhVis21} investigated a range of observational implications and constraints on PBHs with very large masses of $10^{11}-10^{18}M_\odot$. The recent detection of the gravitational waves through pulsar timing in the NANOGrav project has provided additional evidence for the existence of a population of supermassive black holes at high redshifts \citep{Nanograv}.

After the Universe transitions to the matter dominated stage, there is a rapid influx of dark matter onto PBHs. Dark matter forms an extended halo around PBHs. Baryons are initially nearly uniformly mixed with dark matter and also form a halo. The difference in density distribution between baryons and dark matter can be expected only in the vicinity of PBHs, where the energy release from accretion is significant, leading to heating and redistribution of the gas.

The key scale of interest for the Metron telescope is the turnaround radius of the dark matter layer \citep{Ber85}. Its significance can be explained as follows. Initially, dark matter participates in the Hubble expansion and moves away from PBHs. However, the gravitational field of PBHs attracts the layers of dark matter. At a certain point, the velocity of the layer moving away from PBHs becomes zero, and then the layer starts moving towards the PBHs again. During the compression, the layer undergoes mixing and virialization, reducing its size by approximately a factor of two. Over time, more and more distant layers come to a stop and start moving towards the PBHs. The turnaround radius is 
\begin{equation}
r_s(z)=22.4\left(\frac{M_{\rm PBH}}{2\times10^9M_\odot}\right)^{1/3}\left(\frac{1+z}{21}\right)^{-4/3}~\mbox{kpc}.
\label{rseq}
\end{equation}
At redshifts of $z\sim10-20$, the mass of dark matter within this radius is approximately two orders of magnitude greater than the mass of the PBH \citep{DokEro01,DokEro03}. Close to the turnaround radius, strong absorption in the 21 cm line of neutral hydrogen can be expected through the DGE effect \citep{DubrovichGraEro21}. While ionization may occur near the PBH, the degree of ionization decreases at larger distances. Estimates made in \citet{DubrovichGraEro21} showed that at the turnaround radius $r_s$, hydrogen remains mostly neutral, as $r_s$ exceeds the so-called ionization radius. Further investigation into the structure of ``cosmological dinosaurs'' and precise calculations of the ionization radius are supported by the grant from the Russian Science Foundation\footnote{Project 23-22-00013 of the Russian Science Foundation ``Observational Manifestations of Primordial Black Holes in the Distant Universe and in the Solar System'', \url{https://rscf.ru/project/23-22-00013/}.}. From equation (\ref{rseq}), it follows that the angular radius of the absorption ring is approximately 9 arcseconds at $z\sim 20$ and a PBH mass of $2\times10^9M_\odot$.

\begin{figure}
	\begin{center}
\includegraphics[angle=0,width=0.45\textwidth]{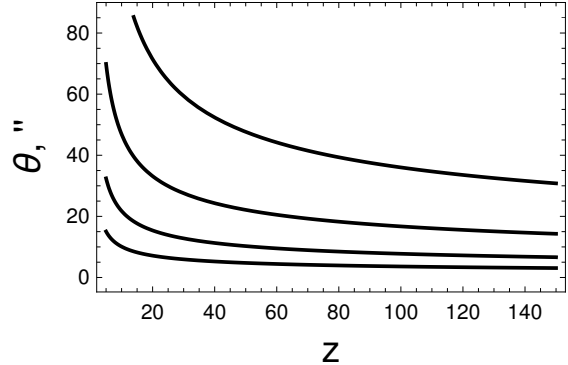}
	\end{center}
\caption{
The angular size of the turnaround radius (\ref{rseq}) in arcseconds as a function of redshift $z$ and PBH masses (from bottom to top) $M_{\rm PBH}=10^9M_\odot$, $10^{10}M_\odot$, $10^{11}M_\odot$, $10^{12}M_\odot$.}
	\label{grangle}
\end{figure}

The angular size of the turnaround radius is shown as a function of the PBH mass and redshift in Figure~\ref{grangle}.
The angular radius of the absorption ring being 9 arcseconds at $z\sim20$ is well within the capabilities of the LOFAR telescope \citep{LOFAR}, but observing the width of the ring would require significantly higher resolution, which will be achievable with the Metron project. Additionally, new computer algorithms for identifying such objects will be necessary, and these are planned to be developed within the framework of the proposed project.

\begin{figure}
	\begin{center}
\includegraphics[angle=0,width=0.45\textwidth]{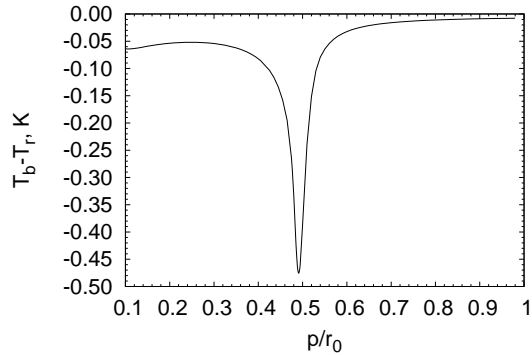}
	\end{center}
\caption{Distribution of brightness across the disk at frequency $\nu=\nu_0$ ($r_0=2r_s=44.8$~kpc).}
	\label{grachgr1}
\end{figure}

In \citet{DubrovichGraEro21}, the spectral distribution and spatial variation of brightness in the 21 cm atomic hydrogen line were calculated using radiation transfer theory. The calculations showed that narrow and deep absorption arises in the form of a spherical shell around PBHs. Here, we present the calculation of the brightness temperature distribution in the 21 cm line for the shell. As an example, we calculate the brightness temperature distributions in frequency and direction for the shell at a redshift of $z=20$.
The central ``hot'' part of the shell with a radius of 5 kpc is excluded from consideration. The outer radius is chosen as $r_0 = 44.8$ kpc, which corresponds to twice the turnaround radius, $r_s$. At $r=r_0$, the conditions in the shell are close to those in an undisturbed medium. The shell is immersed in the background radiation field (which in the modern era is known as the cosmic microwave background radiation). Figure~\ref{grachgr1} shows the distribution of brightness temperature across the disk at the central frequency of the line, $\nu=\nu_0$.

\begin{figure}
	\begin{center}
\includegraphics[angle=0,width=0.45\textwidth]{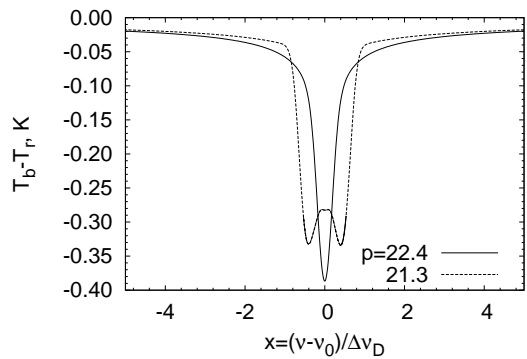}
	\end{center}
\caption{
Frequency profiles of the line for different line-of-sight distances $p=21.3$ and $p=r_s=22.4$~kpc.}
	\label{grachgr2}
\end{figure}

The overall weak decline in brightness when moving from the edge of the disk to the center is determined by the increase in matter density and the decrease in spin temperature. This decrease in spin temperature is caused by the increased probability of photon destruction in the line due to the sharp increase in the de-excitation rate coefficient $q_{10}$ of the upper level of the transition with increasing gas temperature when moving from the edge to the center of the shell. This effect, in particular, determines the line depth in the case of a static shell, as shown in our previous work \citet{DubrovichGra19}.
The sharp decrease in brightness in a narrow region near $p/r_0=0.5$ ($p\approx22$ kpc) is caused by strong absorption of radiation along the line of sight at distances from the center close to the turnaround radius $r_s=22.4$ kpc, where the gradient of the radial velocity is zero. According to Figure~\ref{grachgr1}, the ratio of the width of the spatial intensity distribution at half the maximum to the turnaround radius $r_s$ is 0.05. Figure~\ref{grachgr2} shows the frequency profiles of the 21 cm line for two line-of-sight distances. According to Figure~\ref{grachgr2}, the line depth at $p=r_s$ is approximately --0.4 K, and the relative width at the level of --0.2 K is $\Delta\nu/\nu\approx2.6\times10^{-4}$.

It should be noted that the central mass around which a halo of dark matter forms can be not only a primordial compact halo domain, but also other compact massive objects. Examples of such objects can be closed loops of cosmic strings \citep{BerDokEro11,ShlVilLoe12} or other topological defects.


\section{Study of anomalous absorption at the 21 cm wave}

In 2018, the EDGES telescopes detected anomalous absorption in the 21 cm neutral hydrogen line at redshifts $z=15-20$ \citep{BowRogMon18}. The measured depth of the absorption, 500 mK, is approximately twice the theoretical value of 230 mK. The statistical significance of the deviation between the measured and calculated absorption is $3.8\sigma$. While some studies attempt to explain the EDGES result as observational uncertainties, it remains a mystery in radio astronomy. Models have been proposed to explain this result, suggesting cooling of baryons through a new hypothetical interaction with dark matter particles. The mechanism of absorption could be associated with these new physics phenomena or with underestimation of important factors in the standard picture. In any case, further clarification of the EDGES results is of great fundamental interest.

The Metron project will provide a new perspective to investigate the issue of the mentioned anomalous absorption. The high resolution of Metron will help clarify whether the 21 cm absorption is homogeneous on kiloparsec scales or if it is inhomogeneous. If large-scale inhomogeneities are discovered, it will significantly narrow down the range of absorption models and potentially aid in explaining the EDGES result.


\section{Testing new cosmological models}

Although the standard $\Lambda$CDM cosmological model with an inflationary paradigm is currently dominant, it still needs to explain several observational facts, such as the abundance of early galaxies, massive seeds for early supermassive black holes, the Hubble tension problem, the origin of the background of long-wavelength gravitational waves, galaxies with a deficit of dark matter, and more. These problems are complex but not entirely inexplicable within the framework of the standard $\Lambda$CDM model. However, more significant modifications to the $\Lambda$CDM model in various aspects cannot be ruled out. Since the Metron project aims to observe processes in the early Universe that are sensitive to the background cosmology and the composition of the Universe, it can be expected that the data obtained within its framework can shed light on various aspects of the structure of the Universe.

The fact that the history of cosmology is far from complete is evidenced, for example, by the renewed interest in cyclic or oscillating cosmological models and models with a ``bounce'' \citep{BraPet17}. Cyclic models within the conformal paradigm, with the change of cosmological eons, were developed by R. Penrose and others \citep{GurPen13}. This class of models predicts the existence of ring-like structures in the distribution of relic radiation, but such structures have not been reliably detected yet.

The works of N.N. Gorkavyi and his colleagues \citep{GorVas18, GorTyu21} develop cyclic cosmological models and make numerous predictions that can be consistent with many observational data obtained in recent years. One important prediction is the presence of populations of black holes left over from previous episodes of cosmic expansion and survived its contraction. Unlike PBHs, such black holes are referred to as relic black holes. In this model, a population of early massive galaxies and quasars (with seeds from relic black holes) naturally arises, which can explain the NANOGrav signal \citep{Gor22} and the accelerated expansion of the Universe.

Finally, we should mention the series of works by V.A. Rubakov and his colleagues \citep{AgePetRub21, AgePetRub22}, where new cosmological models with ``genesis'' within the framework of Horndeski theories were fundamentally justified and developed. These models solve the strong coupling problem, where the effective Planck mass approaches zero in the asymptotic past. In these models, a bounce can occur, transitioning to the hot stage of the Big Bang either through an intermediate stage of inflation or without inflation.

An important aspect of future cosmological observations could be the manifestation of microphysics on cosmological scales. In inflationary scenarios, such manifestation includes density perturbations and gravitational waves that originated on small scales but were subsequently stretched to the scales of galaxies, galaxy clusters, and superclusters. In new cosmological models, depending on the nature of dark matter and other factors, microphysics can manifest in new and interesting ways. For example, there are (still very preliminary) indications of the wave nature of dark matter obtained from observations of gravitational lensing \citep{Broetal23}.


\section{Stellar and exoplanetary magnetospheres}

Stars are also sources of radio emissions in the meter range. Detecting their radiation is important for understanding processes in their magnetospheres. In particular, dwarf stars exhibit particularly strong flare activity, which can be accompanied by radio bursts.

One of the scientific breakthroughs in astronomy in recent years has been the discovery and observation of thousands of planets orbiting other stars, known as exoplanets. Observing exoplanets holds great significance in our understanding of the origin of life in the universe. However, the capabilities of optical and infrared observations are limited. Studying exoplanets in the meter radio range with high sensitivity would be an important complement to other methods and could potentially provide unique results. It is worth noting that the first detection of radio emissions from an exoplanet (in the system of the star Tau Boötis) was achieved using the LOFAR ground-based telescope, highlighting the potential of radio observations \citep{Turetal21}.

In the meter range, strong emissions from the magnetospheres of exoplanets and possibly artificial radio sources can be expected. The power of radio emissions from exoplanets can be estimated based on data from Jupiter's radio emissions. Jupiter is known to be the second most powerful source of radio emissions in the meter range after the Sun. Radio emissions in the meter range from Jupiter have a burst-like nature, with typical burst durations of 0.1--1 second and peak fluxes of around $10^6$ Jy. For exoplanets in our Galaxy at typical distances of around 1 kpc, a similar signal would be around $10^{-9}$ Jy. Closer exoplanets would have higher fluxes since the flux decreases with the square of the distance ($1/r^2$ law). Additionally, it can be expected that some giant exoplanets have more powerful magnetospheres, resulting in bursts that are several orders of magnitude stronger.


\section{Conclusions}

The Metron telescope, with its planned characteristics, will be capable of observing extended objects such as early protogalaxies and ``Cosmological Dinosaurs'' that may have existed during the Dark Ages, where absorption features with significant depth of line are formed based on the DGE effect. Metron will allow the study of anomalous absorption in the 21 cm neutral hydrogen line and potentially shed light on its nature. Another goal will be to investigate the magnetospheres of stars and exoplanets. With their placement on Earth, Metron telescopes will also enable the study of various aspects of ionospheric physics and meteorological phenomena.

Among the existing radio telescopes, the Low-Frequency Array (LOFAR), operating in the range of 10--240 MHz, is a competitor to our project \citep{Haaetal13,Gasetal23}. In the Metron project, we plan to achieve significantly better angular resolution and sensitivity by creating large antenna arrays operating as interferometers. Among the projects planned for implementation, a direct competitor can be considered the Square Kilometre Array (SKA), specifically its low-frequency subsystem, SKA-low array, covering the range of 50--350 MHz. In the Metron project, we plan to focus solely on the low-frequency (meter) range, which significantly reduces the overall project cost. We aim to create antenna arrays in this range that are comparable to or even more capable than the SKA-low array in terms of their capabilities.

In the following three papers, we will illuminate the technical specifications of the antennas and equipment of the Metron project depending on the placement of interferometric arrays. In the second paper, the telescope design will be described, while the third paper will discuss the hardware aspect. The fourth paper will describe data processing techniques, including noise removal and extraction of useful astrophysical information.

\section*{acknowledgments}
The authors would like to express their gratitude to the reviewers of the paper for their valuable comments and suggestions.

\section*{FUNDING}
The theoretical estimations comprising 50\% of the content of this paper were made with the support of grant RSF № 23-62-10013. 
Cosmological and galactic applications of the Metron project, comprising 50\% of the paper, were formulated within the framework of the state assignment of the Special Astrophysical Observatory of the Russian Academy of Sciences, approved by the Ministry of Science and Higher Education of the Russian Federation. 

\section*{CONFLICT OF INTEREST}
The authors declare no conflicts of interest.

\bibliographystyle{aspb1}
\bibliography{bibliography}

\end{document}

%% file: sao_cmd_author.tex
\def\squareforqed{\hbox{\rlap{$\sqcap$}$\sqcup$}}

\def\sq{\ifmmode\squareforqed\else{\unskip\nobreak\hfil
\penalty50\hskip1em\null\nobreak\hfil\squareforqed
\parfillskip=0pt\finalhyphendemerits=0\endgraf}\fi}

\def\utw{\smash{\rlap{\lower5pt\hbox{$\sim$}}}}

\def\udtw{\smash{\rlap{\lower6pt\hbox{$\approx$}}}}

\def\diameter{{\ifmmode\mathchoice
{\ooalign{\hfil\hbox{$\displaystyle/$}\hfil\crcr
{\hbox{$\displaystyle\mathchar"20D$}}}}
{\ooalign{\hfil\hbox{$\textstyle/$}\hfil\crcr
{\hbox{$\textstyle\mathchar"20D$}}}}
{\ooalign{\hfil\hbox{$\scriptstyle/$}\hfil\crcr
{\hbox{$\scriptstyle\mathchar"20D$}}}}
{\ooalign{\hfil\hbox{$\scriptscriptstyle/$}\hfil\crcr
{\hbox{$\scriptscriptstyle\mathchar"20D$}}}}
\else{\ooalign{\hfil/\hfil\crcr\mathhexbox20D}}%
\fi}}